\documentclass[twocolumn,aps,prl,showpacs,groupedaddress,amsfonts,amssymb,amsmath]{revtex4-2}
\usepackage{braket}
\usepackage{graphicx}
\usepackage{dcolumn}
\usepackage{bm}
\usepackage{pifont}
\usepackage[colorlinks=true, letterpaper=true, pdfstartview=FitV, linkcolor=blue, citecolor=blue, urlcolor=black]{hyperref}
\usepackage{appendix}

\begin{document}

\title{Disorder-Driven Exceptional Points and Concurrent Topological Phase Transitions}

\author{Xiaoyu Cheng,$^{1,2}$ Tiantao Qu,$^{3}$ Yaqing Yang,$^{1}$ Jun Chen$^{3,4,*}$ and Lei Zhang,$^{1,4,\dagger}$ }
\address{$^1$State Key Laboratory of Quantum Optics Technologies and Devices, Institute of Laser Spectroscopy, Shanxi University, Taiyuan 030006, China\\
$^2$College of Physics and Electronic Engineering, Shanxi Normal University, Taiyuan 030031, China
$^3$State Key Laboratory of Quantum Optics Technologies and Devices, Institute of Theoretical Physics, Shanxi University, Taiyuan 030006, China\\
$^4$Collaborative Innovation Center of Extreme Optics, Shanxi University, Taiyuan 030006, China}

\begin{abstract}
Exceptional points (EPs) are spectral degeneracies unique to non-Hermitian systems, which underpin phenomena from enhanced sensing to unconventional topology. While disorder is usually viewed as detrimental, it can also drive topological phase transitions (TPTs). Here we show that random disorder alone can generate EPs and concurrent TPTs in a multiorbital non-Hermitian lattice with nonreciprocal hopping. Increasing disorder induces successive real-complex-real spectral transitions accompanied by band inversion and quantized changes in the spin Bott index. Using effective medium theory and large-scale simulations, we trace these transitions to a competition between disorder-induced energy-level renormalization and nonreciprocity-driven hybridization. The resulting phase diagram reveals extended EP lines that emerge from the Hermitian TPT point and persist over a broad parameter range. Our results establish disorder as an active mechanism for engineering exceptional point mediated topology in non-Hermitian matter.
	
\end{abstract}

\maketitle
\section{I. INTRODUCTION}
Non-Hermitian systems, characterized by gain and loss or nonreciprocal hopping, have emerged as a fertile platform for discovering spectral and topological phenomena without Hermitian counterparts \cite{bender1998real, heiss2004exceptional, ruter2010observation, regensburger2012parity, zhen2015spawning, feng2017non,miri2019exceptional,ryu2024exceptional,rahmani2024exceptional,guo2009observation,lu2015pt,gong2018topological,shen2018topological,yao2018edge,song2019non,kawabata2019symmetry,bergholtz2021exceptional,ding2022non,okuma2023non,yao2018non,li2020critical, zhang2021observation, liu2021non,zou2021observation,lin2023topological,kawabata2023entanglement,molignini2023anomalous,okuma2020topological}. These systems host exceptional points (EPs) \cite{bender1998real, heiss2004exceptional, ruter2010observation, regensburger2012parity, zhen2015spawning, feng2017non,miri2019exceptional,ryu2024exceptional,rahmani2024exceptional}, where eigenvalues and eigenvectors coalesce, leading to singular responses and unconventional band topology. Over the past decade, EPs have been shown to underpin effects ranging from unidirectional transport and enhanced sensing to the breakdown of bulk-edge correspondence. Parallel developments in non-Hermitian band theory have revealed distinct topological classifications and the non-Hermitian skin effect \cite{yao2018non,li2020critical, zhang2021observation, liu2021non, zou2021observation,lin2023topological,kawabata2023entanglement,molignini2023anomalous,okuma2020topological}, fundamentally reshaping our understanding of states and their localization.

At the same time, disorder viewed as detrimental factor has taken on a new role in topological physics. Random disorder can induce topological phase transitions (TPTs) and even create new topological states absent in clean systems, as exemplified by the topological Anderson insulator \cite{li2009topological,groth2009theory,guo2010topological,liu2017disorder,Li2020Topological,stutzer2018photonic,meier2018observation,cui2022photonic,chen2024realization,zhang2019topological,zhang2020non,gu2023observation,zhang2021experimental,lin2022observation,chen2017disorder,agarwala2019topological,mitchell2018amorphous,yang2019topological,wang2022structural,cheng2023topological,zhou2020photonic,wang2021structural,corbae2023observation,zhang2023anomalous,xing2011topological,qu2024topological,shiACS,quPRB2024,zuo2024}. In both Hermitian and non-Hermitian settings, disorder modifies spectra and topological invariants in ways that cannot be captured by conventional band theory, introducing rich physics governed by effective non-Hermitian self-energies, and Fu and Simon \textit{et.al} pointed out that disorder can give rise to the emergence of EPs or nodal lines in momentum space \cite{papaj2019nodal,zyuzin2019disorder}. Despite these advances, the interplay between random disorder and EPs remains largely unexplored. In particular, it is unknown whether random disorder can generate EPs intrinsically—without requiring engineered gain or loss—and whether such EPs can occur concurrently with disorder-driven TPTs. Resolving this question is crucial for understanding how non-Hermiticity and randomness jointly reshape topology and spectral singularities in realistic systems.

Here, we demonstrate that random disorder can robustly induce EPs and concurrent TPTs in a generic non-Hermitian lattice model with nonreciprocal hopping. While EPs are typically regarded as intrinsic to clean, non-Hermitian Hamiltonians, our results uncover a disorder-enabled route to their formation—sometimes multiple times within the same system. As the disorder strength increases, the spectrum undergoes real-complex-real transitions accompanied by band inversion and a quantized change in the spin Bott index, signaling a topological switch. Using effective medium theory within the self-consistent Born approximation and corroborating large-scale numerical simulations, we reveal that these transitions arise from a competition between disorder-induced energy-level renormalization and nonreciprocity-driven interorbital hybridization. The resulting phase diagram exhibits robust EP lines that emanate from the Hermitian topological transition point and evolve systematically with increasing non-Hermiticity. These findings establish disorder as an active driver of non-Hermitian topology—transforming it from a source of decoherence into a powerful control knob for exceptional-point physics. Our results open a general framework for engineering EP-mediated topological transitions in realistic platforms such as photonic lattices, topolectrical circuits, and cold-atom systems, where both disorder and non-reciprocity are intrinsic or tunable.

\begin{figure}
	\centering
	\includegraphics[width=0.97\linewidth]{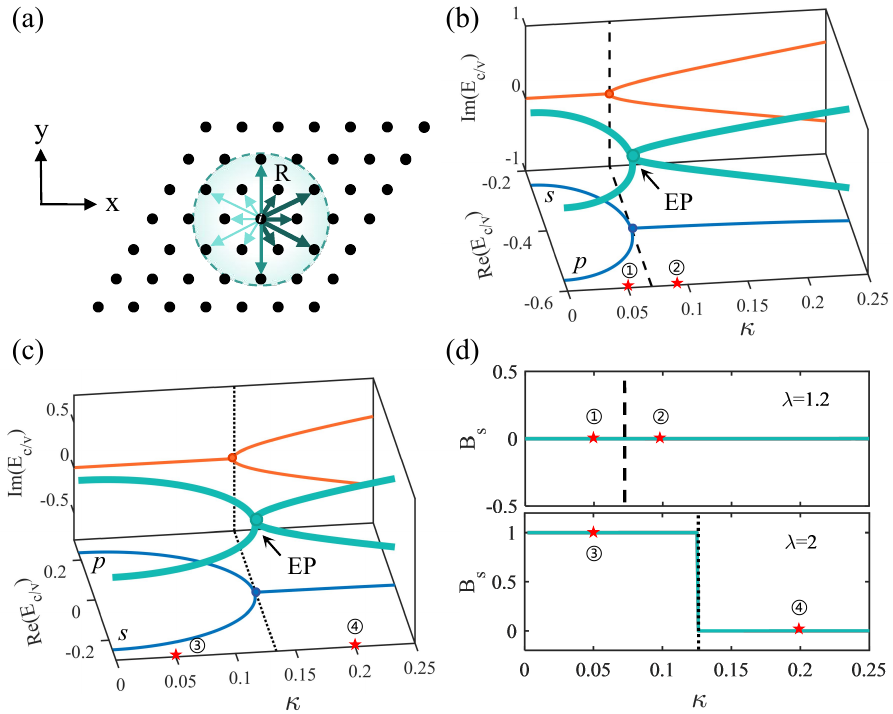}
	\caption{Exceptional points induced by nonreciprocal hopping in a two-dimensional lattice. (a) Schematic of a two-dimensional lattice model with three orbitals per site. Arrows indicate the couplings between atom $i$ and its neighboring atoms, with dark (light) green arrows representing hoppings to the right (left) of atom $i$. The cutoff radius is $R=1.9a$, where $a$ is the lattice constant. (b) and (c) Real and imaginary parts of the conduction ($E_c$) and valence ($E_v$) band energies as a function of non-reciprocal coupling strength $\kappa$ when $\lambda = 1.2$ and $2$, respectively. The other parameters are $a=1$, $\varepsilon_s = 1.8$, $\varepsilon_p= -6.5$, $V_{ss\sigma} = -0.256$, $V_{sp\sigma} = 0.576$, $V_{pp\sigma} = 1.152$, and  $V_{pp\pi}= 0.032$. (d) Spin Bott index $\mathrm{B_s}$ as a function of $\kappa$, with the top and bottom panels shows corresponding to the results in (b) and (c), respectively. All calculations are performed on a $50\times50$ lattice with periodic boundary conditions.}
	\label{Fig1}
\end{figure}

\section{II. NON-HERMITIAN MODEL AND EXCEPTIONAL POINTS IN THE CLEAN LIMIT}
To explore the interplay between disorder, non-Hermiticity and topology, we begin with a generic tight-binding model in a basis with three orbitals  per site \cite{wang2016quantum,huang2018quantum,wang2022structural,Cheng2026035201}. The Hamiltonian in real space is expressed as
\begin{equation}
	\begin{split}
		H&=\sum_{i,\alpha}\varepsilon_{\alpha}\ket{i,\alpha}\bra{i,\alpha}\sigma_0\\
		&+\sum_{\langle i,\alpha;j,\beta \rangle}t_{\alpha,\beta}(\mathbf{r}_{ij})e^{\mathrm{sgn}(x_j-x_i)\kappa }\ket{j,\beta}\bra{i,\alpha}\sigma_0\\
		&+i\lambda \sum_{i} (\ket{i,p_y}\bra{i,p_x}-\ket{i,p_x}\bra{i,p_y} )\sigma_z,\label{HamH}
	\end{split}
\end{equation}
where $i$ and $j$ label lattice sites; $\alpha,\beta \in \{s,p_x,p_y\}$ are orbital indices. ${\varepsilon }_{\alpha }$ describes the on-site energy of the $\alpha$ orbital. $\lambda$ is the spin-orbit coupling strength, $\sigma_{0}$ is the identity matrix, and $\sigma_{z}$ is the Pauli matrix. The hopping integral $t_{\alpha,\beta}(\mathbf{r}_{ij})$ between orbitals $\alpha$ and $\beta$ at sites $i$ and $j$, separated by $\mathbf{r}_{ij}=\mathbf{r}_{j}-\mathbf{r}_{i}$ is defined as \cite{wang2022structural,cheng2023topological}
\begin{equation}
	t_{\alpha,\beta}(\mathbf{r}_{ij}) =
	\frac{\Theta(R-r_{ij})}{r_{ij}^{2}}\rm{SK}
	\begin{bmatrix}
		V_{\alpha\beta\delta},\hat{\mathbf{r}}_{ij}
	\end{bmatrix},\label{HamT}
\end{equation}
where $R$ is the cutoff radius and $\Theta(R-r_{ij})$ is a step function. The function $\rm{SK}[\cdot]$ represents Slater-Koster parametrization for three orbitals, where $V_{\alpha\beta\delta}$ are the bond parameters and $\hat{\mathbf{r}}_{ij}$ denotes the unit direction vector \cite{slater1954simplified}. 
To incorporate non-reciprocity, we introduce the non-reciprocal factor $\kappa$ to the hopping terms. Taking atom $i$ as the reference point, as illustrated in Fig. \ref{Fig1}(a), the factor $e^{\mathrm{sgn}(x_j-x_i)\kappa }$ assigns $e^{+\kappa }$ to the ``rightward'' direction ($x_{j}-x_{i}>0$) within the cutoff radius, while hopping to the ``leftward'' direction ($x_{j}-x_{i}<0$) is weighted by $e^{-\kappa}$. 
The model thus captures a minimal setting where orbital hybridization and non-Hermitian effects coexist.

In the Hermitian limit ($\kappa=0$), the system supports either a topologically trivial or nontrivial (quantum spin Hall-like) phase depending on $\lambda$ \cite{wang2022structural, cheng2023topological}. Introducing nonreciprocal hopping ($\kappa>0$) leads to a collapse of the real energy gap between conduction and valence bands. As shown in Figs. \ref{Fig1}(b) and \ref{Fig1}(c), the real parts of the band energies converge and coalesce at a critical $\kappa$, marking an EP. Beyond this point, the eigenvalues acquire complex-conjugate components, indicating a transition from a real to complex spectrum.

Importantly, this EP also coincides with a TPT when the underlying Hermitian system is initially topologically nontrivial, as evidenced by a sharp change in the spin Bott index [Fig. \ref{Fig1}(d), bottom panel]. Thus, nonreciprocity alone can drive an EP-mediated collapse of topology, while systems starting in the trivial phase experience spectral degeneracy without topological change [Fig. \ref{Fig1}(d), upper  panel]. These results establish the clean non-Hermitian lattice as the foundation for studying how disorder reshapes exceptional-point physics.

\begin{figure}
	\centering
	\includegraphics[width=0.97\linewidth]{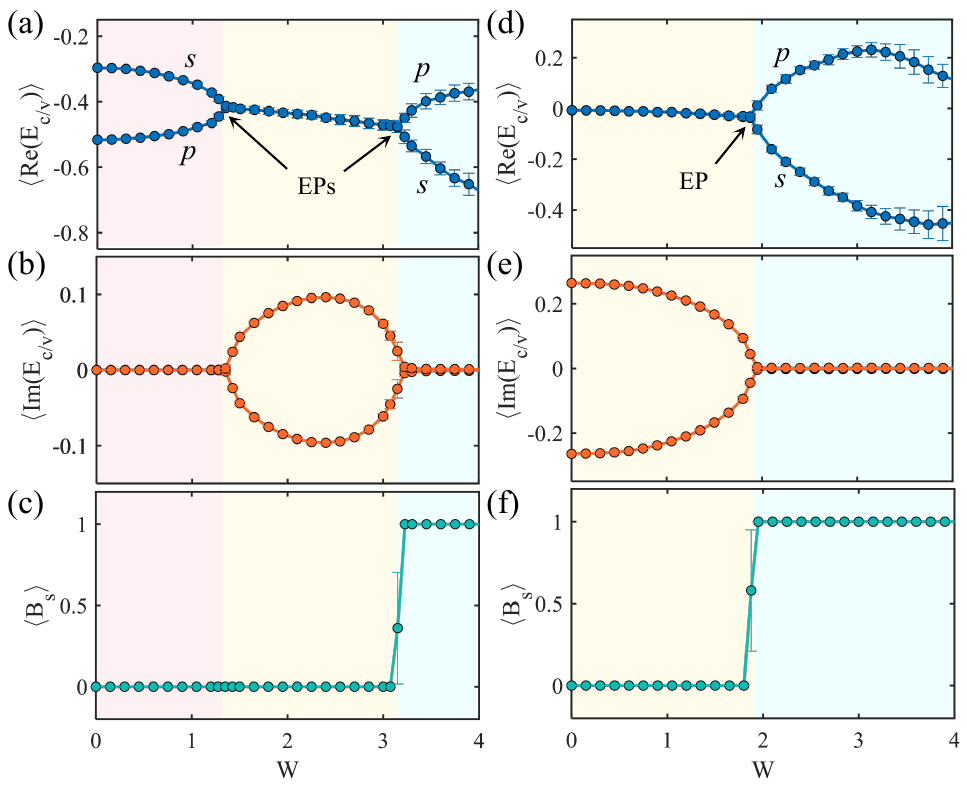}
	\caption{Disorder-driven EPs and TPTs. (a)-(c) Disorder-averaged real parts $\left\langle {\mathrm{Re}(E_{c/v})} \right\rangle$ and imaginary parts $\left\langle {\mathrm{Im}(E_{c/v})} \right\rangle$ and the spin Bott index $\left\langle {\mathrm{B_{s}}} \right\rangle$ as a functions of disorder strength \textsc{w} for $\kappa=0.05$ and $\lambda=1.2$, respectively. (d)-(f) Corresponding results for $\kappa=0.2$ and $\lambda=2$, respectively. Error bars indicate fluctuations across $100$ independent disorder realizations. All other parameters are identical to those in Fig. \ref{Fig1}. The results highlight the emergence of EPs and concurrent TPTs driven by increasing disorder.}
	\label{Fig2}
\end{figure}
\begin{figure}[t]
	\centering
	\includegraphics[width=0.97\linewidth]{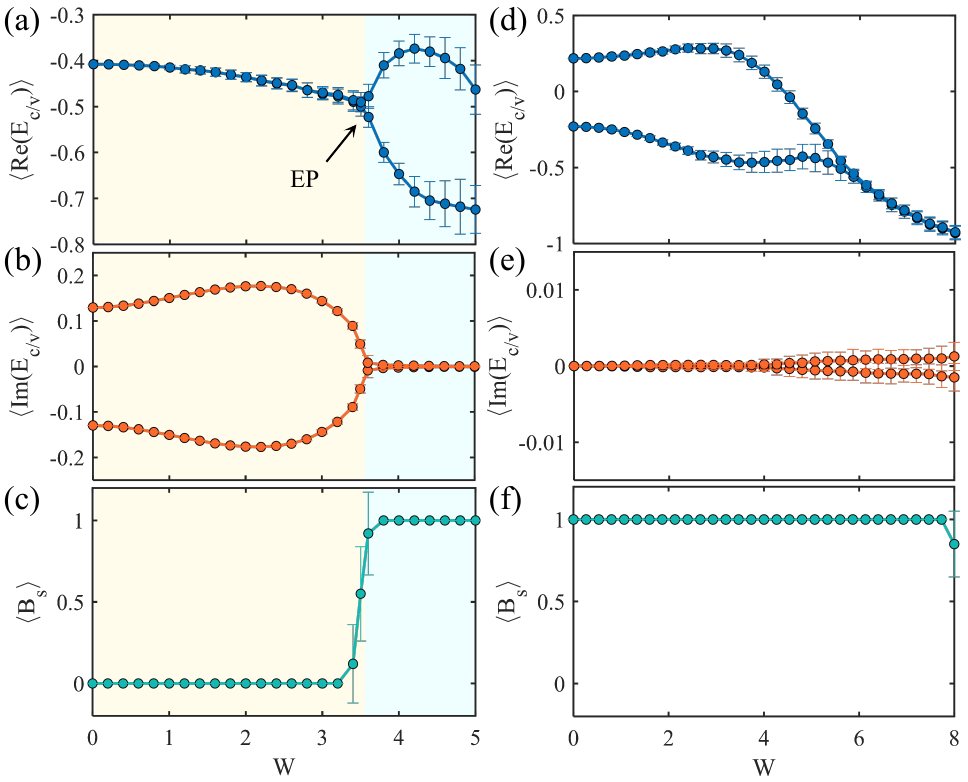}
	\caption{Two complementary cases of disorder-driven EPs and TPTs. (a)-(c) Disorder-averaged real parts $\left\langle {\mathrm{Re}(E_{c/v})} \right\rangle$ and imaginary parts $\left\langle {\mathrm{Im}(E_{c/v})} \right\rangle$ and the spin Bott index $\left\langle {\mathrm{B_{s}}} \right\rangle$ as a functions of disorder strength \textsc{w} for $\kappa=0.09$ and $\lambda=1.2$, respectively. (d)-(f) Corresponding results for $\kappa=0.05$ and $\lambda=2$, respectively. Error bars indicate fluctuations across $100$ independent disorder realizations. All other parameters are identical to those in Fig. \ref{Fig1}. The results highlight the emergence of EPs and concurrent TPTs driven by increasing disorder.}
	\label{Fig5}
\end{figure}

\section{III. DISORDER-DRIVEN EPs AND TPTs}
We now introduce Anderson-type disorder through random on-site potentials, and there is no spatial correlation between different sites, which is given by $\langle V_{i,\alpha} V_{j,\alpha} \rangle = W^2 \delta_{ij}$. $V_{i,\alpha}$ is drawn independently from the uniform distribution $V_{i,\alpha} \in [-\textsc{w}/{2},\textsc{w}/{2}]$, representing static uncorrelated fluctuations in realistic materials. This enables us to examine whether random disorder can itself generate EPs and drive topological transitions in an otherwise trivial non-Hermitian system.

Figures \ref{Fig2}(a)–\ref{Fig2}(c) show the evolution of disorder-averaged band energies and the spin Bott index for a representative trivial system [$\kappa=0.05,\lambda=1.2$, clean case \ding{172} in Fig. \ref{Fig1}(b)]. As the disorder strength $\textsc{w}$ increases, the real parts of the conduction [${\mathrm{Re}(E_{c})}$] and valence [${\mathrm{Re}(E_{v})}$] energies first converge and coalesce, signaling the emergence of a disorder-induced EP. The corresponding imaginary parts become finite and form complex-conjugate pairs, confirming a real-complex transition. Upon a further increase $\textsc{w}$, the spectrum reenters a real regime at a second critical point—a second EP—where the spin Bott index abruptly switches from 0 to 1, marking a disorder-induced TPT. This sequence real$\rightarrow$complex$\rightarrow$real demonstrates that disorder can create EPs, turning randomness into a knob for controlling non-Hermitian spectral topology. Analysis of the orbital character near the second EP (as detailed in Appendix~B) reveals an inversion between $s-$ and $p-$ like components, confirming the topological nature of the transition.

A complementary behavior arises for initially nontrivial systems with stronger nonreciprocity [$\kappa=0.2,\lambda=2$; clean case \ding{175} in Fig. \ref{Fig1}(c)]. The results are presented in Figs. \ref{Fig2}(d)–\ref{Fig2}(f). Here, disorder first preserves a complex spectrum over a broad range, then restores real eigenvalues at a critical $\textsc{w}$, where a single EP coincides with a transition back into a topologically nontrivial phase. In both scenarios, the disorder-induced EPs serve as organizing points for topological change, underscoring their universality.

\begin{figure*}
	\centering
	\includegraphics[width=0.85\linewidth]{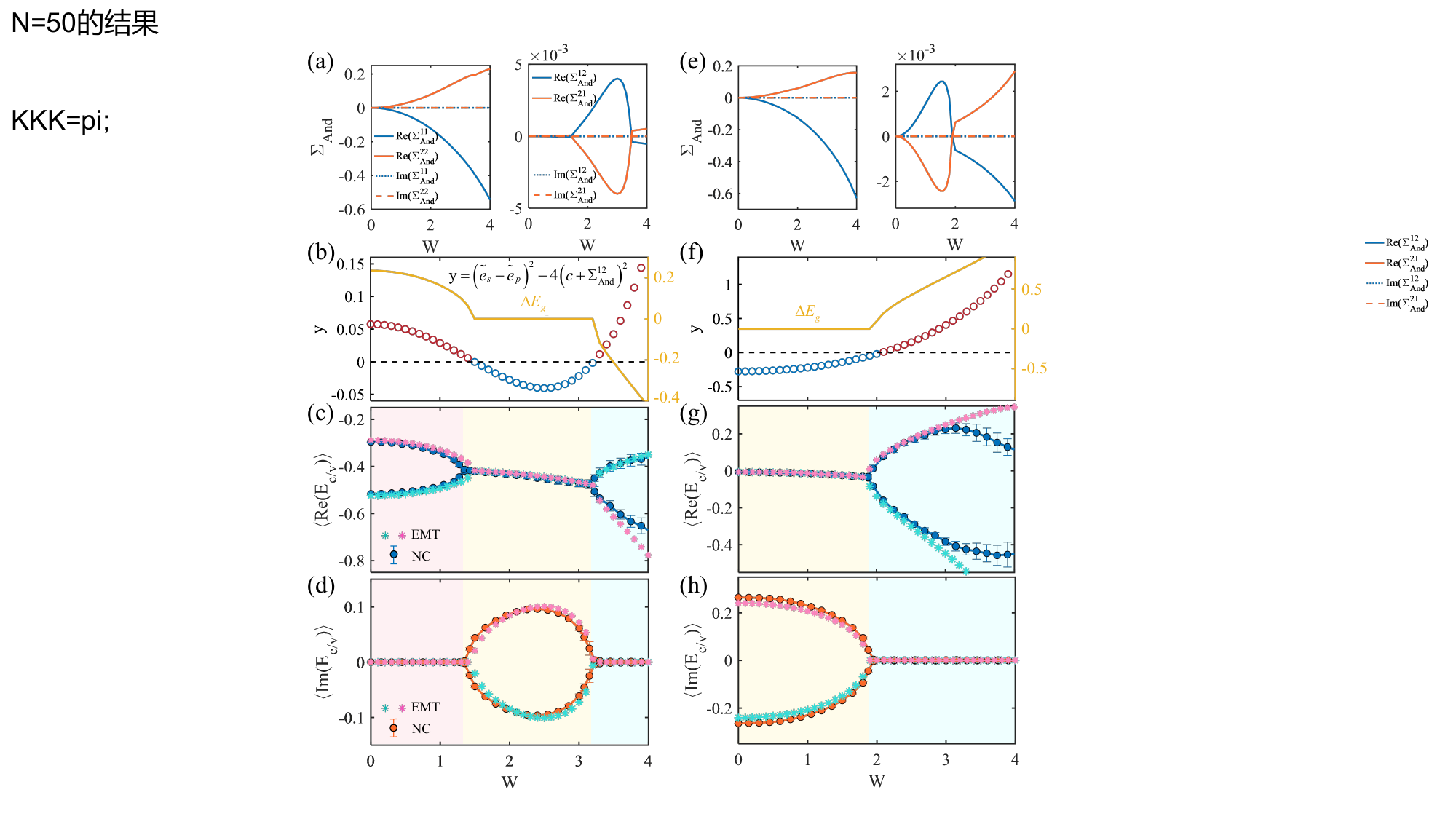}
	\caption{Effective medium theory captures the emergence of disorder-induced EPs and TPTs. (a) and (e) Disordered-averaged self-energy components $\Sigma^{ij}_{\rm{And}}$ as a function of disorder strength \textsc{w} for $\lambda=1.2$ (case \ding{172}) and $\lambda=2$ (case \ding{175}), respectively. Diagonal terms renormalize orbital energies, while the off-diagonal term $\Sigma^{12}_{\rm{And}}$ induces inter-orbital coupling. (b) and (f) Evolution of the discriminant $y$ (left axis) and the energy gap $\Delta E_g$ (right axis) as a functions of \textsc{w}, showing the appearance of EPs (where $y=0$) and gap closing/reopening. For $\lambda=1.2$ a band inversion occurs at the second EP, indicating a TPT.  In contrast, for $\lambda=2$, a TPT also takes place at the EP, marked by the reopening of a real bandgap. (c),(d),(g) and (h) Comparison between effective medium theory (EMT) and real-space numerical calculations (NC). (c) and (d) correspond to $\lambda=1.2$, while (g) and (h) correspond to $\lambda=2$. EMT accurately reproduces the emergence of EPs, real-complex-real spectral transitions, and the associated topological behavior across disorder regimes.}
	\label{Fig3}
\end{figure*}

We further examine two complementary scenarios. A clean non-Hermitian system that is topologically trivial beyond an EP [$\kappa=0.09,\lambda=1.2$, clean case \ding{173} in Fig. \ref{Fig1}(b)]. As shown in Figs. \ref{Fig5}(a)–\ref{Fig5}(c), due to non-reciprocal coupling, the real part of the energy spectrum is degenerate. As the disorder strength increases, the imaginary parts of the spectrum vanish at $W = 3.5$, while the real parts split and a bandgap opens. This behavior indicates the emergence of an EP as the spectrum transitions from complex to real. Notably, the EP coincides with a discontinuous jump in the disorder-averaged spin Bott index $\langle B_s \rangle$ from 0 to 1, confirming that disorder drives the system through an EP, accompanied by a TPT. Another case is a system that is initially topologically nontrivial before reaching an EP [$\kappa=0.05,\lambda=2$, clean case \ding{174} in Fig. \ref{Fig1}(c)]. As shown in Fig. \ref{Fig5}(d), increasing the disorder strength gradually closes the energy gap. When the disorder strength exceeds 4, a prominent imaginary component appears in the spectrum, as shown in Fig. \ref{Fig5}(e), signaling significant spectral broadening. No clear EP is observed in this regime. At sufficiently strong disorder, the system undergoes a TPT (from a topologically nontrivial to a trivial phase). These results highlight that the impact of disorder on non-Hermitian topology crucially depends on the initial phase of the system.

\section{IV. EFFECTIVE MEDIUM THEORY: COMPETITION BETWEEN RENORMALIZATION AND HYBRIDIZATION}
To disentangle the competing effects of energy renormalization and hybridization under disorder, we develop an effective medium theory (EMT) that treats disorder self-consistently. Given that the spin-up and spin-down sectors of the Hamiltonian are decoupled, we focus on analyzing the spin-up subspace. EMT incorporates the effects of disorder through a self-energy correction to the clean Hamiltonian, providing a tractable framework for capturing the essential physics \cite{sheng2006introduction, groth2009theory}. After applying a Fourier transform and band folding, we reduce the problem to a two-band model in momentum space

\begin{equation}
	H_{0}(\vec{k})=
	\begin{pmatrix}
		H_{0}^{11}(\vec{k}) & H_{0}^{12}(\vec{k})\\
		H_{0}^{21}(\vec{k}) & H_{0}^{22}(\vec{k})
	\end{pmatrix}.
\end{equation}
The explicit expressions for the matrix elements $H_{0}^{ij}$ and the full derivation of the EMT framework are provided in Appendix~
C.
Within the self-consistent Born approximation \cite{sheng2006introduction,groth2009theory}, the self-energy due to Anderson disorder is given by
\begin{equation}
	\Sigma_{\rm{And}}=\frac{W^{2}}{48 \pi^{2}}\int_{BZ}d\vec{k}[E+i0^{+}-H_{0}(\vec{k})-\Sigma_{\rm{And}}]^{-1},\label{sceq}
\end{equation}
where the integration runs over the first Brillouin zone. After we self-consistently solve Eq. (\ref{sceq}), the numerical results of self-energy for cases \ding{172} and \ding{175} are presented in Figs. \ref{Fig3}(a) and \ref{Fig3}(e). Notably, the imaginary parts of $\Sigma_{\rm{And}}$ remain vanishingly small, indicating negligible energy-level broadening in the studied disorder regime. This finding agrees with the minimal fluctuations observed in Fig. \ref{Fig2}, where error bars show that the eigenvalue spread is small. We also find that the off-diagonal terms satisfy $\Sigma _{\rm{And}}^{12}=-\Sigma _{\rm{And}}^{21}$, reflecting the underlying orbital coupling symmetry. 

We next diagonalize the effective medium Hamiltonian $H_{\rm{EMH}}=H_{0}+\Sigma_{\rm{And}}$ at the $\Gamma$ point to obtain the disorder averaged conduction and valence eigenvalues
\begin{equation}
	{\langle{E}_{c/v}\rangle}=\frac{{{\widetilde{e}}_{s}}+{{\widetilde{e}}_{p}}\pm \sqrt{{{\left( {{\widetilde{e}}_{s}}-{{\widetilde{e}}_{p}} \right)}^{2}}-4\left( c +\Sigma _{\rm{And}}^{12} \right)^{2} }}{2},\label{eqesep}
\end{equation}
where ${{\widetilde{e}}_{s}}={{\varepsilon }_{s}}+2V_{ss\sigma}(1+3\cosh(\kappa))+\Sigma _{\rm{And}}^{11}$, ${{\widetilde{e}}_{p}}={{\varepsilon }_{p}}+(V_{pp\sigma}+V_{pp\pi})(1+3\cosh(\kappa))+\lambda +\Sigma _{\rm{And}}^{22}$, and $c={\frac{3\sqrt{6}+2\sqrt{2}}{3}{V}_{sp\sigma }\sinh(\kappa)}$. Physically, the diagonal components $\Sigma_{\rm{And}}^{11}$ and $\Sigma_{\rm{And}}^{22}$ correspond to energy-level renormalization and they shift the orbital on-site energies and tune the detuning ${{\widetilde{e}}_{s}}-{{\widetilde{e}}_{p}}$. Increasing disorder gradually drives the two levels toward resonance. The off-diagonal component $\Sigma_{\rm{And}}^{12}$ describes inter-orbital hybridization, arising from nonreciprocal hopping that mixes $s$ and $p$ orbitals. In the clean limit ($\Sigma_{\rm{And}}=0$), the equation under  the square root vanishes when ${{\left( {{\widetilde{e}}_{s}}-{{\widetilde{e}}_{p}} \right)}^{2}}-4 c ^{2}=0 $. At this “critical” $\kappa$, the real part of the gap $ \Delta E_g=\mathrm{Re}({E}_{c}-{E}_{v})$ closes, signaling an EP. Moreover, a TPT coincides with this EP if the underlying Hermitian band gap is inverted $ {{\widetilde{e}}_{s}}<{{\widetilde{e}}_{p}}$, as illustrated in Fig. \ref{Fig1}(c).

In the following, we demonstrate how random disorder drives EPs and concurrent TPTs based on EMT. We begin by analyzing case \ding{172}, where ${{\left( {{\widetilde{e}}_{s}}-{{\widetilde{e}}_{p}} \right)}^{2}}-4c^{2}>0$ and ${{\widetilde{e}}_{s}}>{{\widetilde{e}}_{p}}$ in the clean limit. As shown in Fig. \ref{Fig3}(a), the diagonal self-energy term $\Sigma_{\rm{And}}^{22}$ is positive, raising the valance band ${\widetilde{e}}_{p}$, while $\Sigma_{\rm{And}}^{11}$ is negative, lowering the conduction band ${\widetilde{e}}_{s}$ as disorder strength increases. To illustrate the balance between these two effects—energy-level renormalization tending to close the gap and hybridization tending to reopen it——we introduce the spectral discriminant
\begin{equation}
	y={{\left( {{\widetilde{e}}_{s}}-{{\widetilde{e}}_{p}} \right)}^{2}-4(c +\Sigma_{\rm{And}}^{12})}^{2},\label{disy}
\end{equation} 
which corresponds to the discriminant under the square root in Eq. (\ref{eqesep}), plotted in Fig. \ref{Fig3}(b). Starting from a positive value at zero disorder, $y$ decreases monotonically until it vanishes at the first EP, where the eigenvalues coalesce and the effective Hamiltonian $H_{\rm{EMH}}$ becomes defective. Beyond this point, $y$ turns negative, leading to complex-conjugate eigenvalues with degenerate real parts. With a further increase in disorder, $y$ attains a minimum and then increases back to zero, marking a second EP and restoring purely real eigenvalues, thus completing a characteristic real–complex–real transition. Notably, the second EP coincides with a band inversion [the yellow line in Fig. \ref{Fig3}(b)], signaling a disorder-driven TPT. This nonmonotonic behavior of $y$ reflects the subtle competition between disorder-driven energy renormalization and nonreciprocal hopping mediated inter-level coupling. Importantly, EMT without adjustable parameters quantitatively captures the emergence of both EPs and the concomitant TPT, in excellent agreement with real-space numerical simulations [Figs. \ref{Fig3}(c) and \ref{Fig3}(d)].

In contrast, case \ding{175}, defined by ${{\left( {{\widetilde{e}}_{s}}-{{\widetilde{e}}_{p}} \right)}^{2}}-4c^{2}<0$ and ${{\widetilde{e}}_{s}}<{{\widetilde{e}}_{p}}$ in the clean limit, shows qualitatively distinct behavior. While the signs of the diagonal self-energy components remain consistent with case \ding{172}, no band inversion occurs with increasing disorder. Figure \ref{Fig3}(f) tracks the evolution of $y$, which starts negative and increases to zero at an EP. Beyond this point, $y$ becomes positive, restoring purely real eigenvalues and reopening a real energy gap, as indicated by $\Delta E_g$ [Fig. \ref{Fig3}(f)]. This confirms the occurrence of a TPT coincident with the EP. As before, the EMT predictions closely match numerical results [Figs. \ref{Fig3}(g) and \ref{Fig3}(h)], validating the theoretical framework across different disorder regimes.

\begin{figure*}
	\centering
	\includegraphics[width=0.89\linewidth]{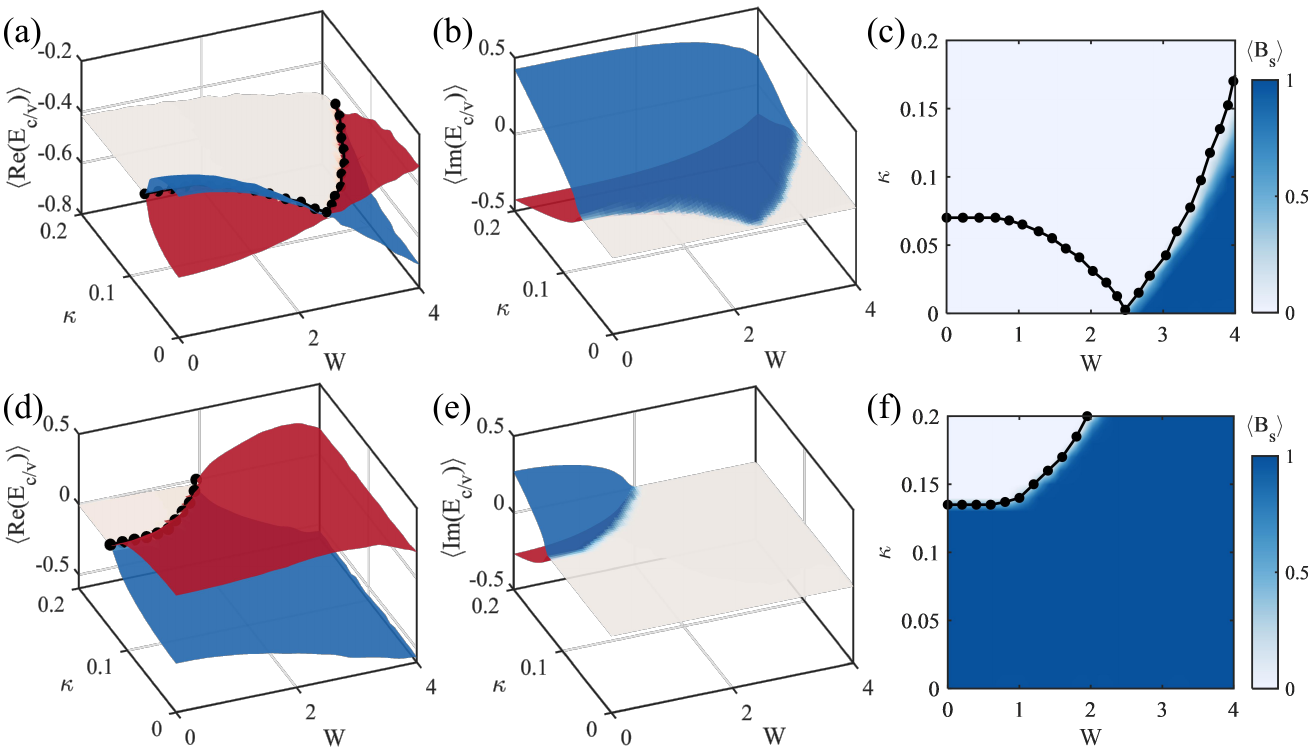}
	\caption{Phase diagram of the energy spectrum and topological invariant in the presence of disorder and nonreciprocity. The disorder-averaged real and imaginary parts of the valence and conduction band energies, $\left\langle {\mathrm{Re}(E_{c/v})} \right\rangle$ and $\left\langle {\mathrm{Im}(E_{c/v})} \right\rangle$, and the spin Bott index $\left\langle {\mathrm{B_{s}}} \right\rangle$ are shown in the parameter space defined by nonreciprocal hopping strength $\kappa$ and disorder strength \textsc{w}. (a)-(c) correspond to case \ding{172}, while (d)-(f) correspond to case \ding{175}. Each data point is averaged over 100 independent disorder configurations. Black dotted lines denote the location of EP lines. All calculations are performed on a $20 \times 20$ with periodic boundary conditions.}
	\label{Fig4}
\end{figure*}

\section{V. PHASE DIAGRAM AND UNIVERSALITY}
Finally, we map the disorder-averaged real and imaginary parts of the conduction and valence band energies, $\left\langle {\mathrm{Re}(E_{c/v})} \right\rangle$ and $\left\langle {\mathrm{Im}(E_{c/v})} \right\rangle$, and the spin Bott index $\left\langle {\mathrm{B_{s}}} \right\rangle$ across the parameter space spanned by nonreciprocal hopping strength $\kappa$ and Anderson disorder strength \textsc{w}. For case \ding{172}, two EP lines (black dotted lines) emerge from the TPT point of the Hermitian case at $\kappa=0$ and $\textsc{w}=2.5$ (see Fig. \ref{supp_k0} in Appendix~D), as shown in Figs. \ref{Fig4}(a) and \ref{Fig4}(b). As $\kappa$ increases, two EP lines separate, with one shifting toward smaller \textsc{w}, while the other, which coincides with the TPT, extends to larger \textsc{w}, as seen in Fig. \ref{Fig4}(c). In contrast, case \ding{175} displays a single EP line, which remains aligned with the TPT throughout the parameter space [Figs. \ref{Fig4}(d)-\ref{Fig4}(f)]. 

The persistence of these EP lines across broad parameter ranges demonstrates that disorder-driven EPs are not fine-tuned accidents, but rather robust features of non-Hermitian topological systems. Together with the EMT analysis, the phase diagram establishes a unified picture: Disorder acts as a self-consistent driving field that renormalizes orbital energies until they reach resonance, at which point non-reciprocal coupling enforces eigenstate coalescence and topological reconfiguration.

\section{VI. CONCLUSION}
Our study revealed a fundamental mechanism by which random disorder can generate and control EPs and drive concurrent TPTs in non-Hermitian lattice systems. Through a combination of large-scale simulations and EMT, we identified a disorder-enabled route to exceptional-point formation that does not rely on fine-tuned gain or loss. This mechanism arises from the competition between disorder-induced energy-level renormalization and nonreciprocity-driven interorbital hybridization, which together govern a robust sequence of real-complex-real spectral transitions. At the EP, these two effects exactly balance, leading to eigenstate coalescence and the onset of a topological switch marked by a quantized change in the spin Bott index.

The EMT framework provides a transparent physical picture: As disorder increases, diagonal self-energies shift orbital levels toward resonance, while off-diagonal components induced by non-reciprocal hopping reopen the hybridization gap. The alternation between these competing tendencies naturally produces two EPs that delimit a regime of complex spectra. The excellent agreement between analytical and numerical results demonstrates that this disorder-induced mechanism is universal and robust, extending well beyond the specific model studied here.

The resulting phase diagram—spanned by disorder strength and non-reciprocity—reveals extended EP lines that emanate from the Hermitian TPT point and persist across a broad range of parameters. This structure unifies previously distinct regimes of Hermitian and non-Hermitian topology and shows that disorder is not merely detrimental and can actively induce and tune exceptional-point topology. Such disorder-enabled EPs represent a new paradigm in non-Hermitian physics: randomness as a self-consistent driving field that reshapes both spectral and topological properties.

These findings open new experimental avenues in photonic lattices \cite{song2020two,lin2024observation}, cold-atom setups \cite{topological250402,liang2022dynamic,zhao2025two}, and electrical circuits \cite{liu2021non,zou2021observation,zhang2023electrical}, where controlled disorder and asymmetric couplings can be used to realize and harness EP-mediated TPTs. These effects open routes to disorder-assisted topological control, enhanced sensitivity near EP lines, and robust mode engineering in imperfect or intentionally disordered non-Hermitian devices. Looking forward, our results could motivate several extensions. Incorporating interactions or nonlinearities may lead to many-body or self-induced exceptional phenomena, while structured or correlated disorder could provide deterministic control over EP formation. Extending this framework to three-dimensional systems may reveal higher-order exceptional surfaces and nodal networks. These directions promise to deepen our understanding of how randomness, non-Hermiticity, and topology intertwine to produce new emergent phases of matter.

\begin{acknowledgments}
	\section{acknowledgments}
This work is supported by the National Key R\&D Program of China under Grant No. 2022YFA1404003, the National Natural Science Foundation of China under Grants No. 12474047 and No. 12574340 and the Research Project Supported by Shanxi Scholarship Council of China. We would like to acknowledge helpful discussions with Professor C.-T. Chan, Professor Z.-Q. Zhang, and Dr. R. Zhang from The Hong Kong University of Science and Technology. This research was partially conducted using the High Performance Computer of Shanxi University.
\end{acknowledgments}

\setcounter{equation}{0}
\renewcommand{\theequation}{A\arabic{equation}}
\appendix
\subsection{APPENDIX A: SPIN BOTT INDEX IN HERMITIAN AND NON-HERMITIAN SYSTEMS \label{SpinBott}} 

\begin{figure*}
	\centering
	\includegraphics[scale=0.75]{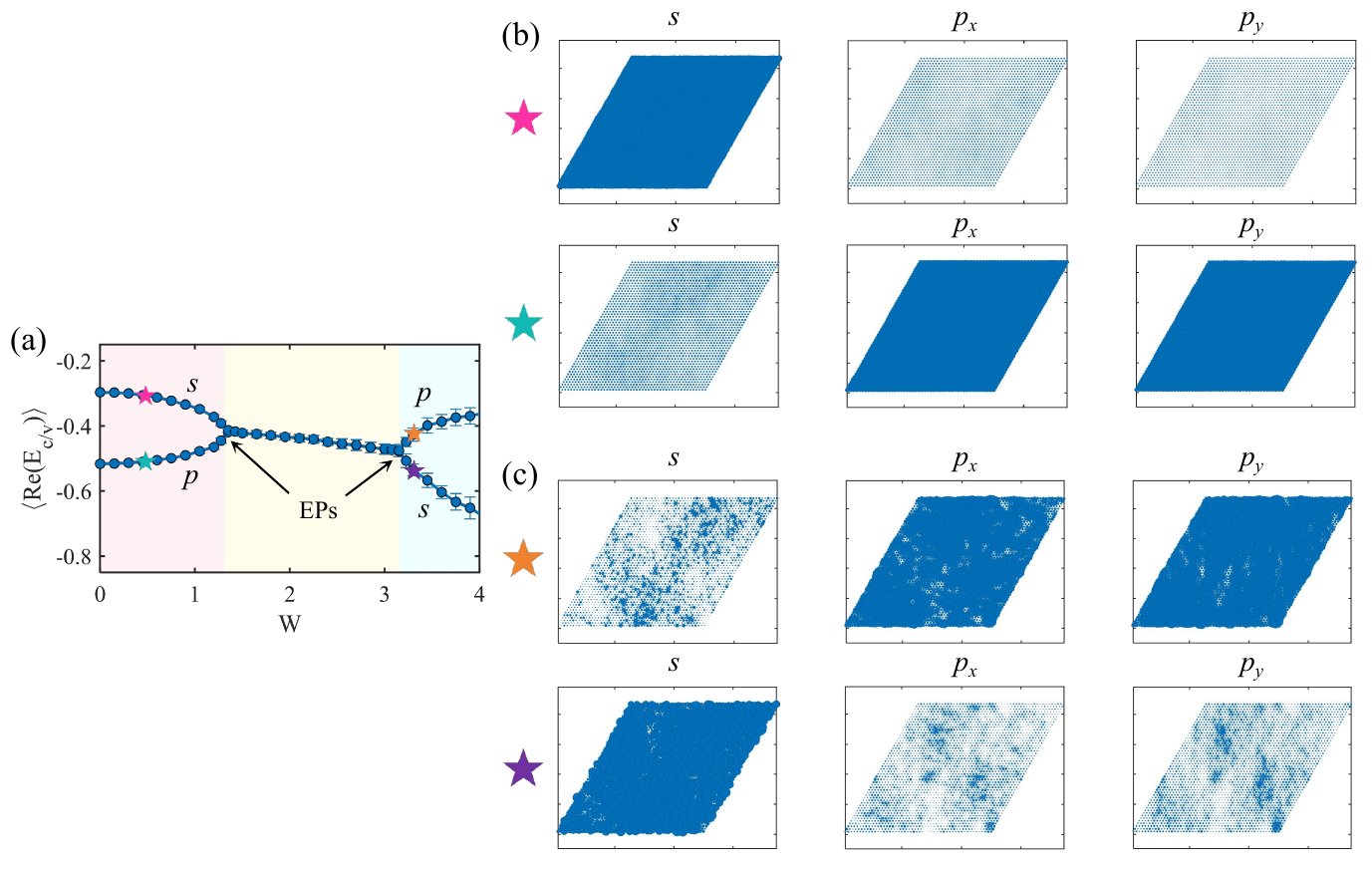}
	\caption{Orbital projection before and after the phase transition point. (a) Disorder-averaged $\langle \mathrm{Re}(E_{c/v}) \rangle$ as a function of disorder strength $W$ for $\kappa = 0.05$ and $\lambda = 1.2$, corresponding to Fig. \ref{Fig2}(a) in the main text. (b) Orbital projection distributions of eigenstates for $W = 0.5$, corresponding to the configuration marked by the pink and cyan stars in (a). (c) Orbital projection distributions of eigenstates for $W = 3.3$, corresponding to a configuration marked by the orange and purple stars in (a).}
	\label{suppfig1}
\end{figure*}

Because Anderson disorder breaks translational symmetry, conventional momentum-space topological invariants (e.g., spin Chern numbers) are inapplicable. Instead, we compute the spin Bott index $B_s$ in real space \cite{huang2018quantum, cheng2023topological}, which remains well-defined for both Hermitian and non-Hermitian systems. In the Hermitian case, we first construct the projector onto occupied states:
\begin{eqnarray}
	P=\sum_{i=1}^{N_{occ}}{|\psi_i\rangle}{\langle\psi_i|},
\end{eqnarray}
where $N_{occ}$ denotes the number of occupied states and $|\psi_{i}\rangle$ is the eigenstate corresponding to eigenvalue $\varepsilon_{i}$. Owing to the conservation of spin $S_z$, the projection operator $P$ naturally decouples into independent spin-up $P_{+}$ and spin-down $P_{-}$ sectors. We then define the projected position operators
\begin{eqnarray}
	U_{\pm}=P_{\pm}\exp(i2\pi X)P_{\pm}+(I-P_{\pm}),\\ V_{\pm}=P_{\pm}\exp(i2\pi Y)P_{\pm}+(I-P_{\pm}),
\end{eqnarray}
where $X$ and $Y$ represent the coordinate operators and $I$ is the identity matrix. To enhance numerical stability, the complementary projectors $Q_{\pm}=I-P_{\pm}$ are incorporated into the projected position operators. A singular value decomposition  $M=Z\varSigma W^{\dagger}$ is then applied to the resulting matrices $U_{\pm}$ and $V_{\pm}$, where $Z$ and $W$ are unitary and $\varSigma$ is a real diagonal matrix.

The Bott indices for two spin sectors are now given by
\begin{eqnarray}
	B_{\pm}=\frac{1}{2\pi}{\rm{Im}}\{{\rm{Tr}}[\log(V_{\pm}U_{\pm}V_{\pm}^{\dagger}U_{\pm}^{\dagger})]\}.
\end{eqnarray}

Then, the spin Bott index $B_s$ is computed as the difference between the Bott indices of the two spin sectors:
\begin{eqnarray}
	B_s=\frac{1}{2}(B_{+}-B_{-}).
\end{eqnarray}

Finally, the disorder-averaged spin Bott index is defined as
\begin{eqnarray}
	\left\langle {B_{s}} \right\rangle=\frac{1}{N}\sum\limits_{i=1}^{N}{{{B}_{s,i}}}.
\end{eqnarray}
where ${B}_{s,i}$ is the spin Bott index of the $i$th disordered realization and the average is taken over $N$ independent configurations.

For non-Hermitian cases, we adopt a biorthogonal basis, $H\left| \psi _{i}^{R} \right\rangle ={{E}_{i}}\left| \psi _{i}^{R} \right\rangle $ and $H^{\dagger}\left| \psi _{i}^{L} \right\rangle =E_{i}^{*}\left| \psi _{i}^{L} \right\rangle$ defining the projection operator as 
\begin{eqnarray}
	P=\sum_{i=1}^{N_{occ}}{|\psi_i^{R}\rangle}{\langle\psi_i^{L}|},
\end{eqnarray}
and proceed with the same steps as above to obtain the spin Bott index $\left\langle {B_{s}} \right\rangle$ in the non-Hermitian situation.

\subsection{APPENDIX B: ORBITAL-RESOLVED BAND CHARACTERISTICS IN DISORDERED SYSTEMS \label{Orbital}} 

Each atom hosts three orbitals, $s,p_{x}$ and $p_{y}$, whose hybridization plays a central role in the emergence of disorder-driven topological phenomena. Figure \ref{suppfig1} shows the orbital-resolved projections of the valence and conduction band eigenstates, evaluated before and after the topological phase transition (TPT) for $\kappa = 0.05$ and $\lambda=1.2$. At weak disorder ($W=0.5$), the conduction band is primarily composed of the $s$ orbital, while the valence band is dominated by the $p$ orbitals. This is consistent with the initial condition ${\varepsilon }_{s}>{\varepsilon }_{p}$, confirming that the system resides in a topologically trivial phase with zero topological invariance. As the disorder strength increases, for instance at $W=3.3$, the orbital character undergoes inversion: The conduction band becomes $p$ dominated, the valence band becomes $s$ dominated, and the topological invariant is equal to 1. In other words, disorder drives the energy band inversion and leads to a TPT. This inversion coincides with a change in the topological invariant from $0$ to $1$, indicating a disorder-induced TPT. This transition is further supported by disorder-induced self-energy corrections $\Sigma_{\rm{And}}$ as shown in Fig. 3(a) in the main text. Specifically, $\Sigma_{\rm{And}}^{11}$ lowers the energy of the $s$ orbital, while $\Sigma_{\rm{And}}^{22}$ raises that of the $p$ orbital, providing a microscopic mechanism for the band inversion and associated TPT.

\begin{figure*}
	\centering
	\includegraphics[scale=0.75]{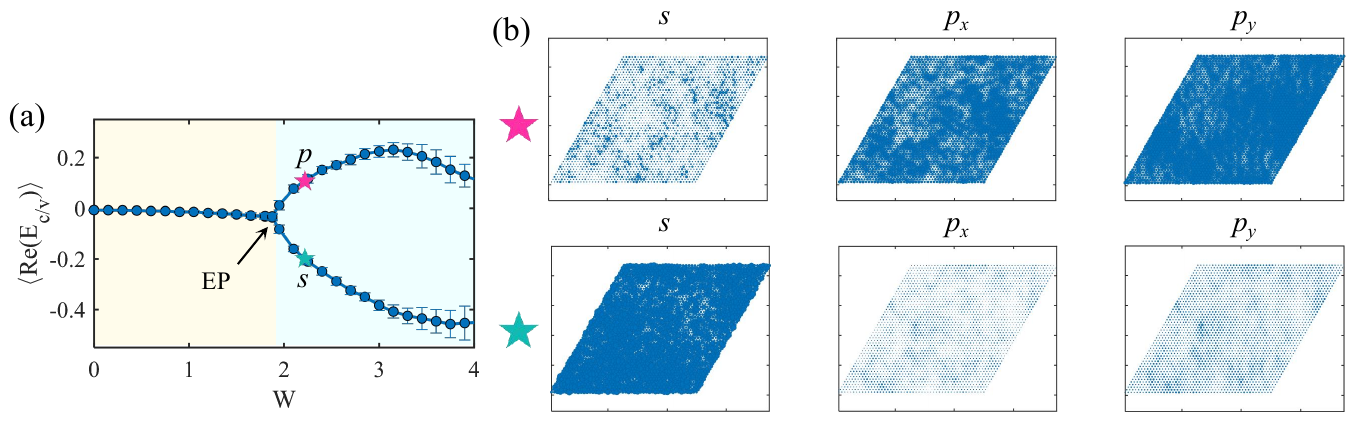}
	\caption{Orbital projection after the phase transition point. (a) Disorder-averaged $\langle \mathrm{Re}(E_{c/v}) \rangle$ as a function of disorder strength $W$ for $\kappa = 0.2$ and $\lambda = 2$, corresponding to Fig. \ref{Fig2}(d) in the main text. (b) Orbital projection distributions of eigenstates at $W = 2.2$, corresponding to the configuration marked by the pink and cyan stars in (a).}
	\label{suppfig2}
\end{figure*}

Figure \ref{suppfig2} presents the orbital-resolved projections of the wave functions for $k = 0.2$ and $\lambda=2$. At disorder strength $W = 2.2$, the conduction band is predominantly composed of $p$ orbitals, while the valence band is mainly $s$ orbital in character. This inversion of orbital content is accompanied by the reopening of the energy gap, signaling a disorder-driven TPT.

\setcounter{equation}{0}
\renewcommand{\theequation}{C\arabic{equation}}

\subsection{APPENDIX C: EFFECTIVE MEDIUM THEORY \label{EMT}} 
To gain a deeper understanding of the interplay between disorder, non-Hermitian effects, and topology, we employ the effective medium theory, which approximates the disorder effect as a self-energy term in the Hamiltonian. This approach simplifies the disordered system while preserving its essential physical properties \cite{groth2009theory, wang2022structural, cheng2023topological}. In this section, we provide a detailed discussion of the calculation of energy bands and TPT 
using effective medium theory.

\begin{widetext}
	First, we apply a Fourier transform to convert the real-space Hamiltonian into momentum space. Since the system conserves spin, the spin-up and spin-down components are degenerate, and we focus on the spin-up component. The Hamiltonian for the spin-up component is then expressed as:
	\begin{equation}
	H=H_{n}+H_{soc}=
	\begin{pmatrix}
		H_{11} & H_{12} & H_{13}\\
		H_{21} & H_{22} & H_{23}\\
		H_{31} & H_{32} & H_{33}\\
	\end{pmatrix}
	+
	\begin{pmatrix}
		0 & 0 & 0\\
		0 & 0 & -i\lambda\\
		0 & i\lambda & 0\\
	\end{pmatrix},
\end{equation}
for non-Hermitian systems, $H\neq H^{\dagger}$ with the matrix elements given by:
\begin{equation}
	H_{11}= \varepsilon_{s}+2V_{ss\sigma}
	\left[2\cos(\frac{\sqrt3}{2}k_{x}-i\kappa)\cos(\frac{1}{2}k_{y})+\cos(k_{y})\right]
	+\frac{2}{3}V_{ss\sigma}
	\left[\cos(k_{x}-i\kappa)+2\cos(\frac{1}{2}k_{x}-i\kappa)\cos(\frac{\sqrt3}{2}k_{y})\right],
\end{equation}
\begin{equation}
	\begin{split}
		H_{22}&=\varepsilon_{p}+(3V_{pp\sigma}+V_{pp\pi})\cos(\frac{\sqrt3}{2}k_{x}-i\kappa)\cos(\frac{1}{2}k_{y}) +2V_{pp\pi}\cos(k_{y})\\
		&+\frac{2}{3}V_{pp\sigma}\cos(k_{x}-i\kappa)+(\frac{1}{3}V_{pp\sigma}+V_{pp\pi})\cos(\frac{1}{2}k_{x}-i\kappa)\cos(\frac{\sqrt3}{2}k_{y}),
	\end{split}
\end{equation}
\begin{equation}
	\begin{split}
		H_{33}&=\varepsilon_{p}+(V_{pp\sigma}+3V_{pp\pi})\cos(\frac{\sqrt3}{2}k_{x}-i\kappa)\cos(\frac{1}{2}k_{y})+2V_{pp\sigma}\cos(k_{y})\\
		&+\frac{2}{3}V_{pp\pi}\cos(k_{x}-i\kappa)+(V_{pp\sigma}+\frac{1}{3}V_{pp\pi})\cos(\frac{1}{2}k_{x}-i\kappa)\cos(\frac{\sqrt3}{2}k_{y}),
	\end{split}
\end{equation}
\begin{equation}
	H_{12}=2\sqrt3iV_{sp\sigma}\sin(\frac{\sqrt3}{2}k_{x}-i\kappa)\cos(\frac{1}{2}k_{y})
	+\frac{2i}{3}V_{sp\sigma}\left[\sin(k_{x}-i\kappa)+\sin(\frac{1}{2}k_{x}-i\kappa)\cos(\frac{\sqrt{3}}{2}k_{y})\right],
\end{equation}
\begin{equation}
	H_{13}=2iV_{sp\sigma}\left[\cos(\frac{\sqrt3}{2}k_{x}-i\kappa)\sin(\frac{1}{2}k_{y})+\sin(k_{y})\right]
	+\frac{2\sqrt{3}i}{3}V_{sp\sigma}\cos(\frac{1}{2}k_{x}-i\kappa)\sin(\frac{\sqrt3}{2}k_{y}),
\end{equation}
\begin{equation}
	H_{23}=\sqrt3(V_{pp\pi}-V_{pp\sigma})\sin(\frac{\sqrt3}{2}k_{x}-i\kappa)\sin(\frac{1}{2}k_{y})
	+\frac{\sqrt3}{3}(V_{pp\pi}-V_{pp\sigma})\sin(\frac{1}{2}k_{x}-i\kappa)\sin(\frac{\sqrt3}{2}k_{y}),
\end{equation}
\begin{equation}
	H_{21}=-2\sqrt3iV_{sp\sigma}\sin(\frac{\sqrt3}{2}k_{x}-i\kappa)\cos(\frac{1}{2}k_{y})
	-\frac{2i}{3}V_{sp\sigma}\left[\sin(k_{x}-i\kappa)+\sin(\frac{1}{2}k_{x}-i\kappa)\cos(\frac{\sqrt{3}}{2}k_{y})\right],
\end{equation}
\begin{equation}
	H_{31}=-2iV_{sp\sigma}\left[\cos(\frac{\sqrt3}{2}k_{x}-i\kappa)\sin(\frac{1}{2}k_{y})+\sin(k_{y})\right]
	-\frac{2\sqrt{3}i}{3}V_{sp\sigma}\cos(\frac{1}{2}k_{x}-i\kappa)\sin(\frac{\sqrt3}{2}k_{y}),
\end{equation}
\begin{equation}
	H_{32}=\sqrt3(V_{pp\pi}-V_{pp\sigma})\sin(\frac{\sqrt3}{2}k_{x}-i\kappa)\sin(\frac{1}{2}k_{y})
	+\frac{\sqrt3}{3}(V_{pp\pi}-V_{pp\sigma})\sin(\frac{1}{2}k_{x}-i\kappa)\sin(\frac{\sqrt3}{2}k_{y}),
\end{equation}
where $\varepsilon_{s}$ and $\varepsilon_{p}$ denote the onsite energies for the $s$ and $p$ orbitals, respectively. The hopping terms $V_{ss\sigma}$, $V_{sp\sigma}$, $V_{pp\sigma}$, and $V_{pp\pi}$ are simplified using the Slater-Koster parametrization method, and $\lambda$ represents the spin-orbit coupling strength. We then transform the basis vectors from $(s, p_{x}, p_{y})$ to $(s, p_{x} + ip_{y}, p_{x} - ip_{y})$, after which the spin-up Hamiltonian becomes:
\begin{equation}
	H=
	\begin{pmatrix}
		H_{11} & \frac{1}{\sqrt2}(H_{12}+iH_{13}) & \frac{1}{\sqrt2}(H_{12}-iH_{13})\\
		\frac{1}{\sqrt2}(H_{21}-iH_{31}) & \frac{1}{2}(H_{22}+iH_{23}-iH_{32}+H_{33})+\lambda & \frac{1}{2}(H_{22}-iH_{23}-iH_{32}-H_{33})\\
		\frac{1}{\sqrt2}(H_{21}+iH_{31}) & \frac{1}{2}(H_{22}+iH_{23}+iH_{32}-H_{33}) & \frac{1}{2}(H_{22}-iH_{23}+iH_{32}+H_{33})-\lambda
	\end{pmatrix}.\label{H}
\end{equation}
To further simplify the problem, we project the three-band Hamiltonian onto a two-band model. Following the procedure outlined in Ref. [\onlinecite{wang2022structural}], we obtain the effective two-band Hamiltonian:
\begin{equation}
	\begin{split}
		H_{\rm{eff}}=&
		\begin{pmatrix}
			H_{11} & \frac{1}{\sqrt2}(H_{12}+iH_{13})\\
			\frac{1}{\sqrt2}(H_{21}-iH_{31}) & \frac{1}{2}(H_{22}+iH_{23}-iH_{32}+H_{33})+\lambda
		\end{pmatrix}+\\
		&\frac{\begin{pmatrix}\frac{1}{\sqrt2}(H_{12}-iH_{13})\\
				\frac{1}{2}(H_{22}-iH_{23}-iH_{32}-H_{33})\end{pmatrix}
			\begin{pmatrix}\frac{1}{\sqrt2}(H_{21}+iH_{31}) & \frac{1}{2}(H_{22}+iH_{23}+iH_{32}-H_{33})\end{pmatrix}}{E-\frac{1}{2}(H_{22}-iH_{23}+iH_{32}+H_{33})+\lambda},
	\end{split}
\end{equation}
where $E$ is the chemical potential.

\begin{figure*}
	\centering
	\includegraphics[scale=0.6]{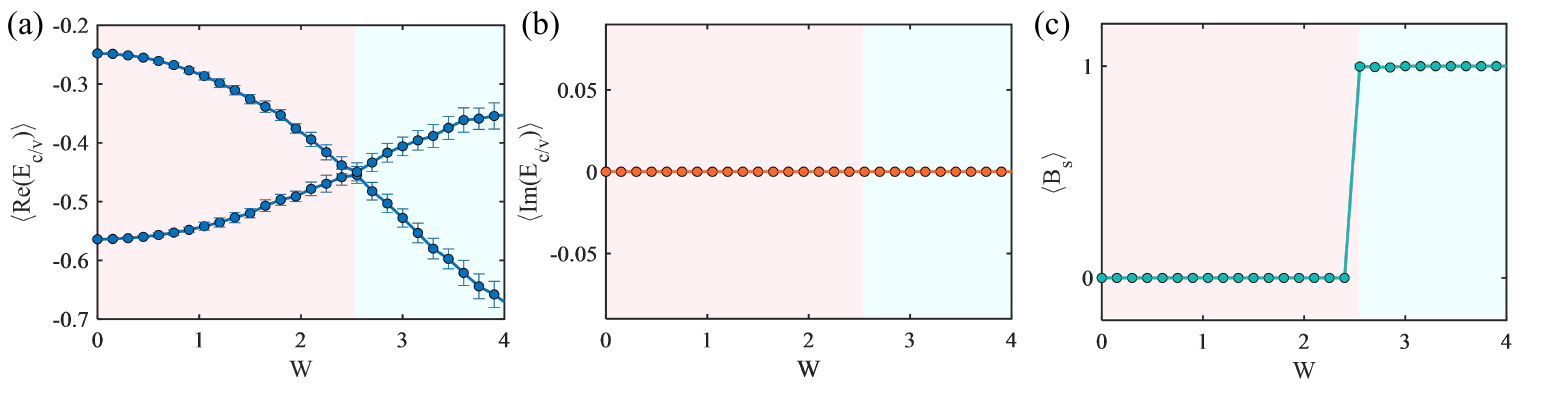}
	\caption{Energy spectrum and topological invariant for Hermitian case. Disorder-averaged (a) $\left\langle {\mathrm{Re}(E_{c/v})} \right\rangle$, (b) $\left\langle {\mathrm{Im}(E_{c/v})} \right\rangle$ and (c) $\left\langle {\mathrm{B_{s}}} \right\rangle$ as function of disorder strength  $W$ in Hermitian system for $\kappa=0$ and $\lambda=1.2$. Each data point is averaged over 100 disorder configurations, and other parameters are the same as Fig. 1.}
	\label{supp_k0}
\end{figure*}
Next, we take the effective Hamiltonian as the starting point, $H_{0} = H_{\rm{eff}}$. Using the self-consistent Born approximation \cite{sheng2006introduction}, we can compute the self-energy $\Sigma_{\rm{And}}$ due to Anderson disorder from the following equation \cite{groth2009theory}:
\begin{equation}
	\Sigma_{\rm{And}}=\frac{W^{2}}{48 \pi^{2}}\int_{BZ}d\vec{k}[E+i0^{+}-H_{0}(\vec{k})-\Sigma_{\rm{And}}]^{-1},
\end{equation}
where $W$ represents the disorder strength and the integral is over the first Brillouin zone. 

Finally, we obtain the expression of the effective medium Hamiltonian as:
\begin{equation}
	\begin{split}	
		H_{\rm{EMH}}&=H_{0}+\Sigma_{\rm{And}}=
		\begin{pmatrix}
			H_{\rm{0}}^{11}+\Sigma_{\rm{And}}^{11} & 	H_{\rm{0}}^{12}+\Sigma_{\rm{And}}^{12}\\
			H_{\rm{0}}^{21}+\Sigma_{\rm{And}}^{21} & 	H_{\rm{0}}^{22}+\Sigma_{\rm{And}}^{22}
		\end{pmatrix}.
	\end{split}
\end{equation}

Since the band inversion occurs at the $\Gamma$ point, we set $k_{x} = k_{y} = 0$, allowing the effective medium Hamiltonian to be written in a more concise form.
\begin{equation}
	H_{\rm{EMH}}=
	\begin{pmatrix}
		\varepsilon_{s}+2V_{ss\sigma}(1+3\cosh(\kappa))+\Sigma_{\rm{And}}^{11} & \frac{3\sqrt{6}+2\sqrt{2}}{3}V_{sp\sigma}\sinh(\kappa)+\Sigma_{\rm{And}}^{12}\\
		-\frac{3\sqrt{6}+2\sqrt{2}}{3}V_{sp\sigma}\sinh(\kappa)+\Sigma_{\rm{And}}^{21} & \varepsilon_{p}+(V_{pp\sigma}+V_{pp\pi})(1+3\cosh(\kappa))+\lambda+\Sigma_{\rm{And}}^{22}
	\end{pmatrix}.
\end{equation}

Based on the numerical results showing the evolution of the self-energy elements with $W$ in Fig. \ref{Fig3}(a) of the main text, we observe that the off-diagonal element satisfies $\Sigma_{\rm{And}}^{21} = -\Sigma_{\rm{And}}^{12}$, leading to the following expression for the general solution:
\begin{equation}
	{\langle{E}_{c/v}\rangle}=\frac{{{\widetilde{e}}_{s}}+{{\widetilde{e}}_{p}}\pm \sqrt{{{\left( {{\widetilde{e}}_{s}}-{{\widetilde{e}}_{p}} \right)}^{2}}-4\left( c +\Sigma _{\rm{And}}^{12} \right)^{2} }}{2},\label{eqesep1}
\end{equation}
where we define the quantities ${{\widetilde{e}}_{s}} = {{\varepsilon }_{s}} + 2V_{ss\sigma}(1 + 3\cosh(\kappa)) + \Sigma_{\rm{And}}^{11}$, ${{\widetilde{e}}_{p}} = {{\varepsilon }_{p}} + (V_{pp\sigma} + V_{pp\pi})(1 + 3\cosh(\kappa)) + \lambda + \Sigma_{\rm{And}}^{22}$, and $c = \frac{3\sqrt{6} + 2\sqrt{2}}{3} V_{sp\sigma} \sinh(\kappa)$. The band gap is given by $\Delta E_{
	g} = \mathrm{Re}(E_{c} - E_{v})$.

\subsection{APPENDIX D: DISORDER-DRIVEN
	TOPOLOGICAL PHASE TRANSITIONS IN THE 
	HERMITIAN SYSTEM \label{Hermitian}} 

In the Hermitian system, as shown in Fig. \ref{supp_k0}, the disorder-averaged $\langle \mathrm{Re}(E_{c/v}) \rangle$ gradually approach each other as the disorder strength $W$ increases. At $W = 2.5$, the disorder-averaged bulk band gap closes, accompanied by a sudden jump in the disorder-averaged spin Bott index $\langle \mathrm{B_s} \rangle$ from 0 to 1. This indicates that, in the Hermitian case, the disorder-driven system undergoes a TPT. The Hermitian nature of the system ensures that all eigenenergies are real.
\end{widetext}
\bigskip

\noindent{$^{*)}$chenjun@sxu.edu.cn}\\
\noindent{$^{\dagger)}$zhanglei@sxu.edu.cn}

\bibliography{ref}

\end{document}